\documentclass{article}
\usepackage{amsmath}
\usepackage{amssymb}
\usepackage{graphicx}

\newtheorem{theorem}{Theorem}
\newtheorem{acknowledgement}[theorem]{Acknowledgement}

\input{tcilatex}
\begin{document}

\title{On Hofstadter butterfly spectrum: Chern-Simons theory, subband gap
mapping, IQHE and FQHE labelling}
\author{F. A. Buot$^{1,2,3}$, G. Maglasang$^{1,3}$, A. R. Elnar$^{1,3}$ and
C. M. Galon$^{1,3}$ \\
%EndAName
$^{1}$CTCMP, \ Cebu Normal University, Cebu City 6000\\
Philippines,\\
$^{2}$C\&LB Research Institute, Carmen, Cebu 6005\\
Philippines,\\
$^{3}$LCFMNN, University of San Carlos, Cebu City 6000,\\
Philippines}
\maketitle

\begin{abstract}
The magnetic field affects the Bloch band structure in a couple of ways.
First it breaks the Bloch band into magnetic subbands or the Landau levels
are broadened into magnetic Bloch bands. The resulting group of subbands in
the central portion of the energy scale is associated with the integer
quantum Hall effect (IQHE). Then at high fields it changes the integrated
density of states of the remaining lowest and topmost subband, respectively,
which can be associated with fractional quantum Hall effect (FQHE).

Here, we employ the Maxwell Chern-Simons gauge theory to formulate the
subband-gap mapping algorithm and to construct the butterfly profile of the
Hofstadter spectrum. The two regions in the spectrum responsible for the
IQHE are identified. At very high magnetic fields the highest and lowest
subband are affected by magnetic-field induced restructuring of the
integrated density of states in each subband, respectively. The resulting
transformation of each of the two subband is responsible for the FQHE. Thus,
in the central regions of the energy scale, the principal group of subbands
defined by the gap mapping is responsible for the IQHE. The fine structure
of the topmost and lowest subband, which convey an iterative nature of the
magnetic spectrum is a result of a hierarchical scaling and restructuring by
the magnetic fields on the integrated density of states in each respective
subband and is responsible for the FQHE.

\pagebreak
\end{abstract}

\section{INTRODUCTION}

When electrons are subjected to both a magnetic field and a periodic
electrostatic potential, two-dimensional systems of electrons exhibit a
complex energy spectrum, known as the Hofstadter's butterfly \cite{hops}.
This complex spectrum results from an incommensurate/commensurate interplay
between two characteristic length scales, defined by the magnetic lattice
with lattice constant given by the magnetic length, $l_{B}$, and the
periodic electrostatic potential with lattice constant denoted by $a$. The
magnetic length $l_{B}$ has the physical meaning of the smallest size of a
circular orbit in a magnetic field which is allowed by the uncertainty
principle. Following Hofstadter \cite{hops} and Wannier \cite{wannier}, we
assume the square lattice model with the two quantizing fields which perhaps
yield the first fractal-like energy spectrum discovered in physics.

\begin{figure}[h!]
\centering
\includegraphics[width=4.9294in]{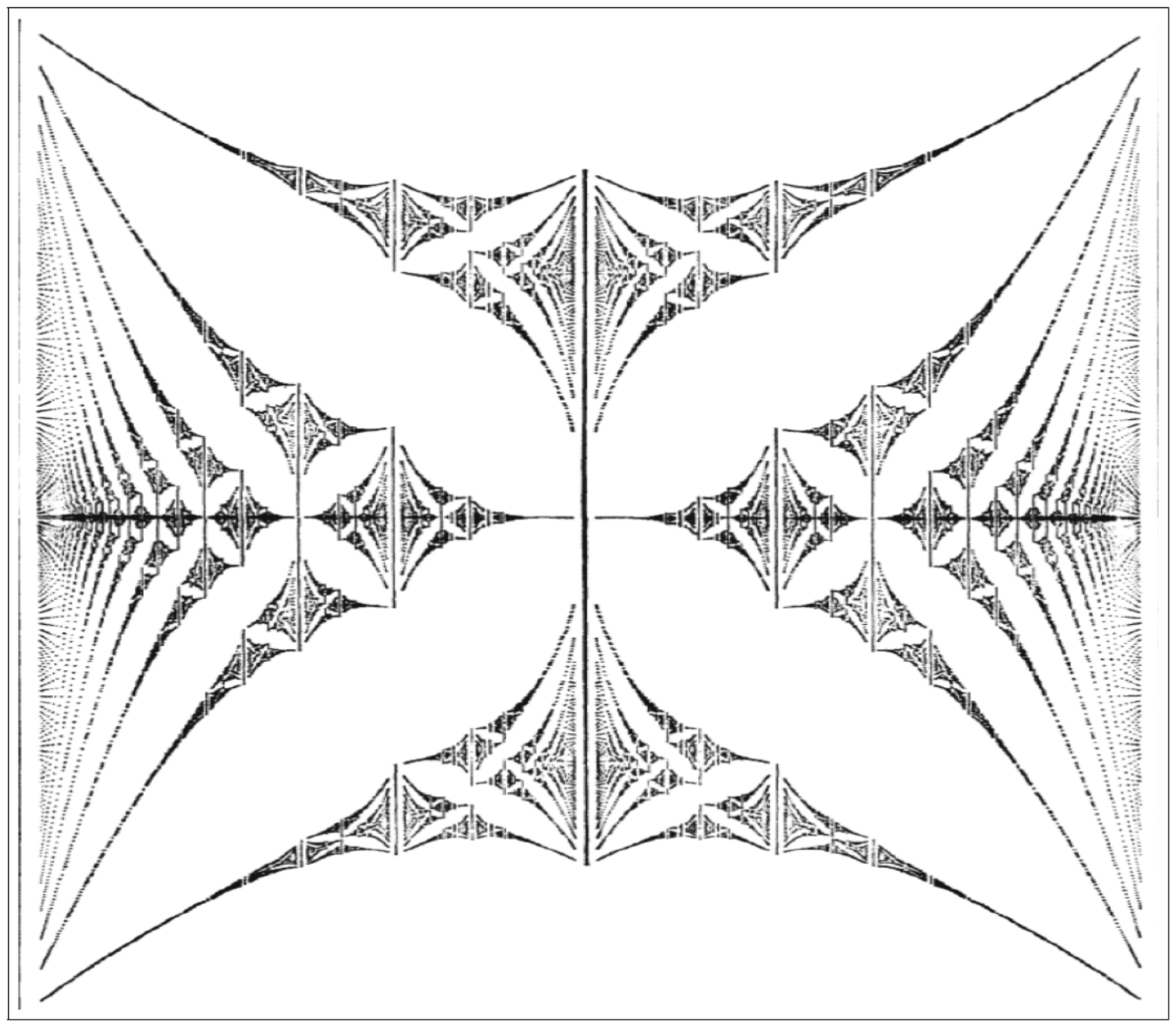}
\caption{The Hofstadter butterfly energy spectrum obtained from computer
graphics rendering of computer numerical simulation results. One
characteristic feature is the wide band gap between topmost subband and
lowest subband. This wide subband gap appears to allow the field to
restructure the separated lower and upper central regions. Indeed, these two
subbands are restructured left and right of the middle of the abscissa.
[Figure taken from Hofstadter \protect\cite{hops}]}
\label{fig1}
\end{figure}

The Schr\"{o}dinger difference equation studied by Hofstadter\cite{hops} is
known as the Harper's equation and can be written as%
\begin{equation}
u\left( m+1\right) +u\left( m-1\right) +2\cos \left( m\left( \frac{2\pi }{n}%
\right) -v\right) u\left( m\right) =\epsilon u\left( m\right)  \label{harper}
\end{equation}%
where $m$ and $n$ are integers. We shall see that $n$ denotes the number of
magnetic subbands in what follows. Obviously, this equation has periodicity
in $\Phi =\frac{2\pi }{n}$. The energy spectrum obtained\cite{hops} is shown
in Fig. \ref{fig1}

To explain the above complex spectrum, Wannier \cite{wannier}, using
counting arguments, derived the so-called Diophantine equation to map out
the energy subband gaps of the spectrum. Here, we propose a simpler counting
argument.

We employ the Maxwell Chern-Simons (CS) gauge theory\cite{bfqhe, maxcs} to
start the counting arguments for subband gap mapping. First we set the basic
variables in analyzing the complex spectrum using CS gauge theory.
Acknowledging the symmetry and periodicity of the complex energy spectrum
shown in Fig. \ref{fig1}, we then draw the lines defining the bottom and top
of the subbands, i.e, the subband gaps. Our results is depicted in Fig. \ref%
{fig2} and Fig. \ref{fig3}. These figures will be discussed in more detail
in what follows.

\begin{figure}[h!]
\centering
\includegraphics[width=5.1906in]{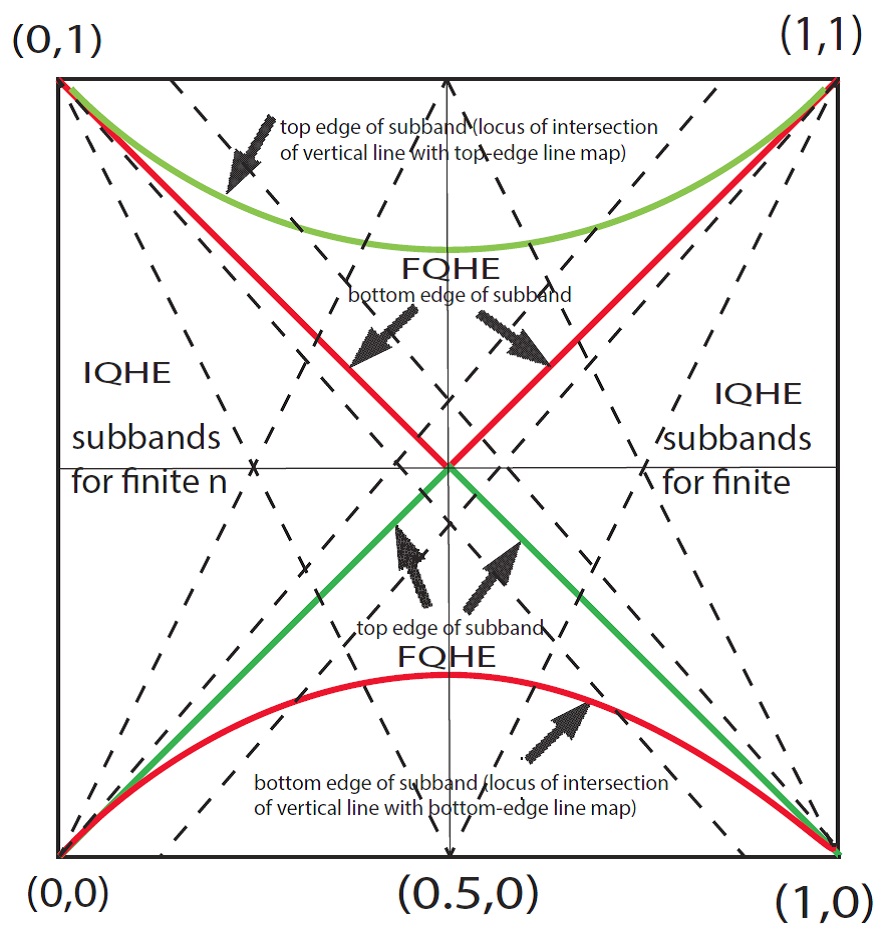}
\caption{The lines in the figure map the center of the subband gaps. There
are essentially four regions of the spectrum, separated by the diagonal
lines. Remarkably, the diagonal lines represent wide subband gaps in Fig.%
\protect\ref{fig1}. The central lower and upper regions of the spectrum is
counted as the first lower subband and the first upper subband,
respectively. These two subbands clearly show up in the middle of the
spectrum. However, to the left and right of the middle, these two subbands
clearly undergo restructuring reminiscent of FQHE. Therefore, we label these
two, lower and upper central regions, of the spectrum as the FQHE regions.}
\label{fig2}
\end{figure}

More of the lines are drawn in Fig.\ref{fig3} to show how the rays, defining
different values or numeration counting of the subbands emanates from the
four corners of the spectrum.

\begin{figure}[h!]
\centering
\includegraphics[width=4.926in]{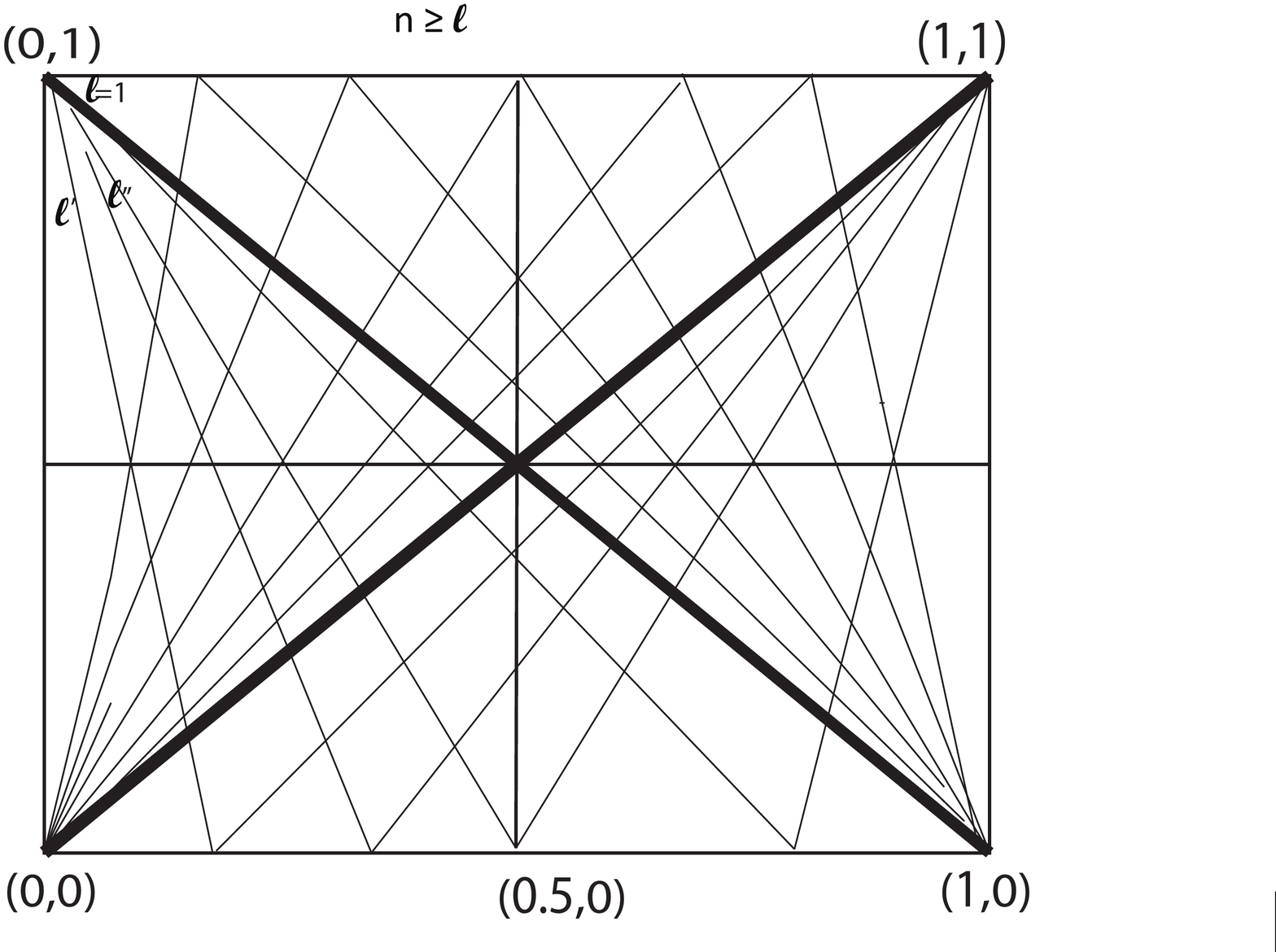}
\caption{The figure shows how the lines emanates from the four corners of
the spectrum.}
\label{fig3}
\end{figure}

\section{MAXWELL CHERN-SIMONS GAUGE THEORIES}

We write the Chern-Simons Lagrangian density for $U\left( 1\right) $ gauge
theory for $2$-dimensional system of manifold, $M$, as%
\begin{equation}
\mathcal{L}_{CS}=\gamma \varepsilon ^{\mu \lambda \nu }A_{\mu }\partial
_{\lambda }A_{\nu }-A_{\mu }J^{\mu }  \label{Lcs}
\end{equation}%
where $J^{\mu }=\left( \rho ,\vec{J}\right) $, $\rho $ is the charge density
and $\vec{J}$ is the current density. Later, we will associate the parameter 
$\gamma $ with our magnetic parameters. Equation (\ref{Lcs}) is often
referred to as the Maxwell Chern-Simons theory \cite{maxcs}. The equation of
motion is obtained by variation with respect to $A_{\mu }$%
\begin{equation*}
\frac{\delta \mathcal{L}_{CS}}{\delta A_{\mu }}=\gamma \varepsilon ^{\lambda
\nu }\partial _{\lambda }A_{\nu }-J^{\mu }=0
\end{equation*}%
This gives%
\begin{equation}
\gamma \int\limits_{M}\nabla \times \vec{A}\cdot
da=\int\limits_{M}J^{0}=\int\limits_{M}\rho  \label{eqmotion2}
\end{equation}%
$\int\limits_{M}\rho $ is equal to the integrated density of states of each
subband. In general, each subband have the same integrated density of
states, reflecting the same degeneracy for each Landau levels in free
electrons case.

We now show that the total density of states is an integer, equal to the
magnetic flux divided by the quantum flux, which is equal to the degeneracy
of a Landau level in a magnetic field. We do this by first counting the
number of states in magnetic phase-space. This is given by \cite{bfqhe}%
\begin{eqnarray*}
\mathcal{N} &=&\frac{1}{2\pi \hbar }\iint \vec{\nabla}\times \mathcal{\vec{K}%
}\cdot d\vec{a} \\
&=&\frac{1}{2\pi \hbar }\iint \vec{\nabla}\times \frac{e}{c}\vec{A}\cdot d%
\vec{a}
\end{eqnarray*}%
where $\mathcal{\vec{K}}=\vec{P}+$ $\frac{e}{c}\vec{A}+\vec{F}ct$, \cite%
{covar} $\Phi $ is the total magnetic flux, and $\vec{F}$ is the uniform
electric field. We will ignore the electric field from now on. Therefore,
the Landau level degeneracy, $\mathcal{N}$, is%
\begin{eqnarray*}
\mathcal{N} &=&\frac{1}{2\pi \hbar }\frac{e}{c}\iint \vec{B}\cdot d\vec{a} \\
&=&\frac{\Phi }{2\pi \left( \frac{\hbar c}{e}\right) }=\frac{\Phi }{\phi _{o}%
}
\end{eqnarray*}%
where $\left( \frac{2\pi \hbar c}{e}\right) =\phi _{o}$ is the quantum flux.
For a square lattice, the number of quantum states in one Bloch band is
equal to the number of lattice sites, $N$. We denote by $n$ the number of
magnetic subband formed from a Bloch band under a uniform magnetic field.
Since the magnetic field does no work or does not impart energy on the
systems, we can assume that the number of quantum states in a Bloch band is
conserved when uniform magnetic field is applied. Thus for $n$ subbands, we
have,%
\begin{eqnarray}
N &=&\frac{n}{2\pi \hbar }\iint \vec{\nabla}\times \frac{e}{c}\vec{A}\cdot d%
\vec{a}=n\frac{Na^{2}}{2\pi \left( l_{B}\right) ^{2}}  \notag \\
&=&n\frac{\Phi }{\phi _{o}}  \label{kfactor2}
\end{eqnarray}%
In other words, 
\begin{equation*}
n=\frac{N}{\left( \frac{\Phi }{\phi _{o}}\right) }=\frac{N}{\mathcal{N}}
\end{equation*}%
where the Landu level (LL) degeneracy $\mathcal{N}$ is, 
\begin{equation*}
\mathcal{N}=\left( \frac{\Phi }{\phi _{o}}\right)
\end{equation*}%
Thus, we can measure the whole Bloch band in units of the LL degeneracy to
give us the number of subbands. We end up with%
\begin{equation}
n\frac{Na^{2}}{2\pi \left( l_{B}\right) ^{2}}=N  \label{normbnd}
\end{equation}%
where $\mathcal{N=}$ $\frac{Na^{2}}{2\pi \left( l_{B}\right) ^{2}}$ is the
degeneracy of each subband, $a$ is the lattice constant of a $2$-D square
lattice and $l_{B}=\sqrt{\frac{\hbar c}{eB}}$. Putting $N=1$ or cancelling
on both sides of Eq. (\ref{normbnd}), we have effectively normalized our
Bloch band width to unity.

We redefine our flux variable as%
\begin{equation*}
\Phi =\frac{a^{2}}{2\pi \left( l_{B}\right) ^{2}}=\frac{1}{n},\text{ \ \ \ \
\ \ }0\leq \frac{1}{n}\leq 1\text{, \ \ }
\end{equation*}%
Then we have%
\begin{equation}
n=\frac{1}{\frac{a^{2}}{2\pi \left( l_{B}\right) ^{2}}}=\frac{1}{\Phi }
\label{units}
\end{equation}%
In Eq. (\ref{units}), the number of subbands is determined in units of $\Phi
.$We see that our normalized Bloch band is now measured in units of $\Phi $.
We see that when $\Phi =1$, the magnetic lattice coincides with the periodic
atomic potential lattice, and hence one expects that a single Bloch band is
restored, $E\left( k\right) $ is periodic in $k$ and in reduced Brillouin
zone or one band, $0\leq k\leq \frac{2\pi }{a}=\frac{2\pi }{l_{B}}$ with $%
l_{B}$ the magnetic length, i.e., consisting of only one subband equal to
the whole Bloch band width.

\section{SUBBAND GAP MAPPING}

In tracing the first gap from the top of the Bloch band width, we reckon
that the \textit{bottom} of the first gap from the \textit{topmost} first
subband must lie at values of one value of $\Phi $ down in the Bloch
energy-band scale. This will define the diagonal line given by the linear
equation,%
\begin{equation*}
W_{bottop}=1-\Phi 
\end{equation*}%
starting from $\left( 0,1\right) $ coordinate, where $W$ is the '$y$%
-coordinate' representing the points in the normalized zero-field Bloch band
width. Other subbands \textit{bottom} boundaries will be defined by lines
given by the linear equation,%
\begin{equation*}
W_{t}=1-t\Phi \text{, \ \ \ \ \ \ }t\leq n\text{, }t\text{ integer}
\end{equation*}%
starting from $\left( 0,1\right) $ coordinate, where $n$ is the total number
of subbands.

On the other hand, \textit{top} boundary of the \textit{lowest} subband must
lie after a one count of $\Phi $. This define another diagonal line of a
linear equation,%
\begin{equation*}
W_{toplow}=\Phi
\end{equation*}%
stating from $\left( 0,0\right) $ coordinate. These diagonal lines are
depicted in Figs. \ref{fig1} and \ref{fig2}. Other subbands top boundaries
will be defined by lines given by the linear equation,%
\begin{equation*}
W_{l}=l\Phi \text{, \ \ \ \ }l\leq n\text{, \ \ \ }l\text{ integer}
\end{equation*}%
stating from $\left( 0,0\right) $ coordinate.

\subsection{Reflection symmetry at $\Phi =\frac{1}{2}$}

\begin{figure}[h!]
\centering
\includegraphics[width=3.384in]{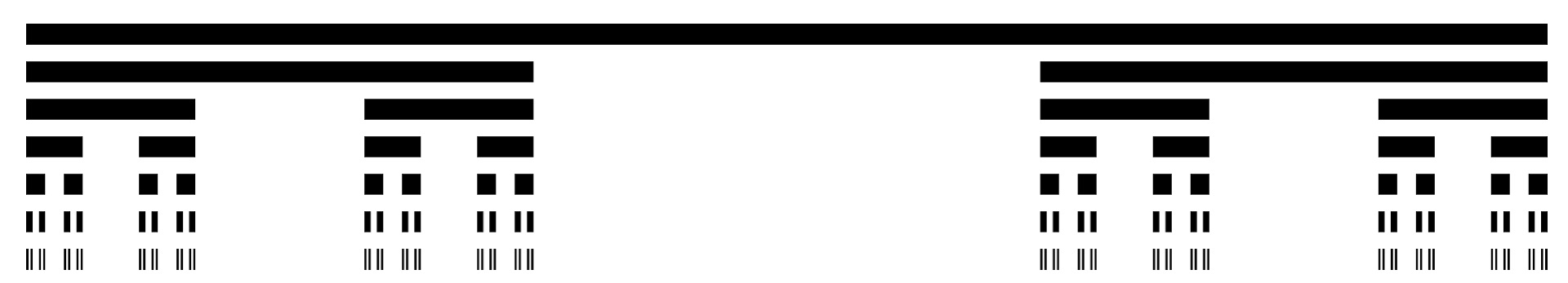}
\caption{An example of the iterative construction of a Canto set. [Taken
from Wikipedia]}
\label{fig4}
\end{figure}

The Hofstadter butterfly spectrum is somewhat reminiscent of a Cantor set,
which is an example of \textit{fractal} string. In Fig.\ref{fig4}, the
horizontal line may represent the zero-field Bloch band width. By
tentatively identifying the Hofstadter butterfly spectrum with the cantor
set, we see that the limiting lowest number of subbands is $n\geq 2$.
Physically of course, the lower integer value $n=1$ just coincide with the
original zero-field Bloch band width. Just like the Cantor set the gap
between the limiting two subbands is the largest in the set. This is also
physically meaningful since at this magnetic field strength the flux is
large, or the degeneracy is large and concentrated in the two subbands.
Moreover, the cyclotron frequency spacing, $\hbar \omega =\frac{eB}{mc}$, is
also large and the two subbands are located symmetrically near the middle of
the zero-field Bloch band width. Since $n=\infty $ and $n=1$ are physically
identical, giving the whole zero-field Bloch band width at this limiting
cases, while also considering the periodicity implied by Eq. (\ref{harper})
of period in$\frac{2\pi }{n}$ $\left( n=\infty \text{ and }n=1\right) $, we
expect reflection symmetry of the integer number of subbands at $\frac{2\pi 
}{n}=\pi $ or $n=2$. We can therefore identify $n=1$ with $n=\infty $, and
trace the gaps back towards $n=2$, with decreasing integer $n$ as a complete
mirror image of the left portion of the flux corresponding to $n=2$. Thus,
although reminiscent of Cantor set this cyclic characteristic, among others,
of the Hofstadter butterfly spectrum then gives a diametrical contrast to
the iterative construction of the Cantor set.

\subsection{IQHE and FQHE Labelling}

One most notable characteristics of the Hofstadter butterfly is the
reconstruction of the topmost subband as well as that of the lowest subband.
Although the left and right portion of the spectrum can be characterize as
regions of IQHE\ regions \cite{previous, comments, jagna}, the top and
bottom regions of the spectrum can be identified with the phenomenon of
FQHE, since obviously this involves the reconstruction of a single subband 
\cite{bfqhe} at high magnetic fields . This reconstruction is evidenced by
the measurements \cite{stormer1, stormer2,stormer3} and Chern-Simons
theoretical explanation of the FQHE given by Buot \cite{bfqhe}. These are
all indicated in Figs. (\ref{fig2}) and (\ref{fig3}).

\section{CONCLUDING REMARKS}

We need to say something about the profile of the \textit{top} boundary of
the \textit{topmost} subband, as well as the \textit{bottom} boundary of the 
\textit{lowest} subband. The bottom profile of the lowest subband boundary
must of the form of a convex curve with endpoints connecting the point $%
\left( 0,0\right) $ and $\left( 1,0\right) $. Similarly, the profile of the
top boundary of the topmost subband must be a concave curve with endpoints
at $\left( 0,1\right) $ and $\left( 1,1\right) $. This the main reason for
the butterfly profile of the spectrum. This is reminiscent of the Landau
levels shifting the lowest energies to higher energies, and the highest
energy to lower energies in the presence of magnetic flux, reflecting the
increase of $\hbar \omega =\frac{eB}{mc}$ without changing the total
energies, since the magnetic field does not work on the system. This aspect
of the spectrum is indicated in Figs. (\ref{fig2}) and (\ref{fig3}).

\begin{acknowledgement}
The author is grateful for the 'Balik Scientist' Visiting Professor grant of
the PCIEERD-DOST, Philippines, at Cebu Normal University, Cebu City,
Philippines.
\end{acknowledgement}


\begin{thebibliography}{99}
\bibitem{hops} D.R. Hofstadter, \textit{Energy levels and wave functions of
Bloch electrons in rational and irrational magnetic fields}, Phys. Rev. 
\textbf{B14}, 2239 (1976).

\bibitem{wannier} G.H. Wannier, \textit{A Result Not Dependent on
Rationality for Bloch Electrons in a Magnetic Field}, Phys. Stat. Sol. (b) 
\textbf{88}, 757 (1978).

\bibitem{bfqhe} F.A Buot, G. Maglasang, A.R.F. Elnar,and C.M. Galon, \textit{%
On Fractional Quantum Hall Effect (FQHE): A Chern-Simons and nonequilibrium
quantum transport Weyl transform approach}, arXiv: 2107.13970 (2021).

\bibitem{maxcs} T. Van Mechelen and Z. Jacob, \textit{Viscous
Maxwell-Chern-Simons theory for topological electromagnetic phases of matter}%
, Phys. Rev. B \textbf{102}, 155425 (2020).

\bibitem{covar} Felix A. Buot, \textit{On quantum Hall effect: Covariant
derivatives, Wilson lines, gauge potentials, lattice Weyl transforms, and
Chern numbers}, arXiv:2106.16238 (2021).

\bibitem{previous} Felix A. Buot, \textit{Nonequilibrium superfield and
lattice Weyl transform approach to quantum Hall effect,} arXiv:2001.06993

\bibitem{comments} Felix A. Buot, \textit{Comments on the Weyl-Wigner
calculus for lattice models},http://arxiv.org/abs/2103.10351

\bibitem{jagna} Felix A. Buot, \textit{On the quantization of Hall effect in
electrical conductivity: A nonequilibrium quantum superfield and lattice
Weyl transform transport approach}, AIP Conference Proceedings \textbf{2286}%
, 030007 (2020).\textit{\ Quantum-Based Devices}, Phys. Rev. B\textbf{42},
9429-9456 (1990).

\bibitem{stormer1} Stormer, H. (1992). \textit{Two-dimensional electron
correlation in high magnetic fields}. Physica B: Condensed Matter, 177(14),
401-408.

\bibitem{stormer2} R. Willett, J. P. Eisenstein, H.L. Stormer, D. C. Tsui,
A. C. Gossard and J. H. English, \textit{Observation of an Even-Denominator
Quantum Number in the Fractional Quantum Hall Effect}, Phys. Rev. Letts. 59,
1776 (1987).

\bibitem{stormer3} W. Pan, H. L. Stormer, D. C. Tsui, L. N. Pfeiffer, K.W.
Baldwin, and K.W.West, \textit{Fractional Quantum Hall Effect of Composite
Fermions}, Phys. Rev. Letts. \textbf{90}, 016801-1 (2003).
\end{thebibliography}
\end{document}